\documentstyle[seceq,letter,psfig]{ptptex}
\newcommand{\beq}{\begin{equation}}
\newcommand{\beqa}{\begin{eqnarray}}
\newcommand{\eeq}{\end{equation}}
\newcommand{\eeqa}{\end{eqnarray}}
\newcommand{\simg}
   {\mathrel{\raise.3ex\hbox{$>$\kern-.75em\lower1ex\hbox{$\sim$}}}}
\newcommand{\siml}
   {\mathrel{\raise.3ex\hbox{$<$\kern-.75em\lower1ex\hbox{$\sim$}}}}

\title{
Lens Model Degeneracy and Cosmological Tests by Strong Gravitational Lensing
}
\author{%       %Use \sc for the family name
Takeshi {\sc Chiba} and Ryuichi {\sc Takahashi}
}

\inst{
Department of Physics, Kyoto University, 
Kyoto 606-8502, Japan
}

\recdate{%      %Editorial Office will fill in this.
}

\abst{
We estimate the sensitivity of lensing observables to the parameters in 
the lens model (isothermal sphere/Navarro-Frenk-White profile) 
parameters and to cosmological parameters. We find that the 
observables are primarily dependent on the lens model parameters, 
while the dependence on cosmological 
parameters is minor (especially so for the dark energy parameters). 
We demonstrate the lens model degeneracy by deriving both the projected mass 
density profile and the circular velocity profile.  
We also identify a possible source of the problem of fitting the averaged 
mass profile of CL 0024+1654 with the Navarro-Frenk-White profile.
}

\begin{document}

\maketitle

%\section{Introduction}
{\bf 1.Introduction.}
Gravitational lensing has been extensively studied as a useful cosmological 
tool to probe the high redshift universe. \cite{sef}
In particular, it has been used to provide  limits on 
cosmological parameters. 
For example, the statistics of gravitational lensing of QSOs by 
intervening galaxies provide a powerful tool to set constraints on the 
cosmological constant.\cite{kochanek} 
Lensed images of distant galaxies in cluster, so called arcs or rings, 
may provide a bound on the cosmological constant or even on the equation 
of state of dark energy.\cite{fh,fy} 
This is mainly because the distance relation 
$D_{LS}/D_S$, where $D_{LS}$ is the distance from the lens to the source and 
$D_S$ is that from the observer to the source, has a strong dependence on the 
cosmological constant. \cite{ffkt}

However, results obtained regarding gravitational lensing also depend 
on the lens model. Hence, in using it as a tool to obtain bounds on 
cosmological parameters, we must be careful about the uncertainties 
inherent in the lens model itself. If derived values of gravitational 
lensing observables are sensitive to the lens model, constraints
obtained on cosmological parameters may have a meaning whose implication 
will only become clear after resolving issues concerning the lens model.
Only recently have these  uncertainties begun to be systematically  
studied for the case of gravitational lens statistics. \cite{lo}\tocite{wts} 
In this letter, we elucidate the problem involving the model
uncertainties by estimating the sensitivity of observables to the
parameters in the lens model and cosmological parameters. 
We consider the giant arc system observed in 
CL 0024+1654 as an example. As an aside, we point out that the NFW 
profile fit may not be excluded by the current data. 

\vspace{10pt}
%\section{Lens model and observables}
{\bf 2.Lens model and observables.}
%We shall derive lens equation and define critical curves for 
% axially symmetric lenses \cite{sef}.
%\subsection{lens equation and critical curves}
Let $\eta$ and $\xi$ be the source position and the impact  
position in the lens plane, respectively. Define a length scale $\xi_0$
in the lens plane and a corresponding length scale 
$\eta_0=\xi_0 D_S/D_{L}$. Here  $D_L$ and $D_S$ are the angular diameter 
distance to the lens and source, respectively. 
Then in terms of the dimensionless quantities $x=\xi/\xi_0$ and 
$y=\eta/\eta_0$, the lens equation is given by
\beq
y=x-{m(x)\over x},
\label{lens}
\eeq
where $m(x)=2\int^x_0 (\Sigma(x') / \Sigma_{cr}) x'dx'$, and
$\Sigma_{cr}= D_S / 4\pi G D_LD_{LS}$.
Here $D_{LS}$ is the angular diameter distance between the lens and 
the source. The determinant of the Jacobian 
$A\equiv \partial {\bf y}/\partial {\bf x}$ of the mapping 
Eq.(\ref{lens}) is calculated as
\beq
{\rm det} A=\left(1-{m\over x^2}\right)\left(1-{d\over dx}
\left({m\over x}\right)\right).
\label{critical}
\eeq
The critical curves for axisymmetric lenses are those for which 
${\rm det} A=0$. Circles satisfying $m/x^2=1$ are called 
``tangential critical curves'', while those satisfying $d(m/x)/dx=1$ are 
called ``radial curves''.

%\subsection{ isothermal model}

For an isothermal model with a core, the mass profile is 
$\rho(r)=\sigma^2 / 2\pi G (r^2+r_c^2)$,
 with $\sigma$ being the one-dimensional velocity dispersion and 
$r_c$ the core radius. 
The radius of the tangential critical curve, $\theta_E$, is then given by
\beq
\sqrt{\theta_E^2+\theta_c^2}+\theta_c=4\pi\sigma^2{D_{LS}\over D_S},
\label{th:iso}
\eeq
where $\theta_c=r_c/D_L$. 
The circular velocity at radius $r$ is
\beq
v^2(r)={GM(\leq r)\over r}=2\sigma^2\left(1-
{r_c\over r}\tan^{-1}{r\over r_c}\right).
\label{vel:iso}
\eeq

%\subsection{NFW model}
%\paragraph*{NFW model}

For the Navarro-Frenk-White (NFW) model,\cite{nfw} the mass profile 
is given by $\rho(r)=\rho_s\left(r/r_s\right)^{-1}
\left(\left(r/r_s\right)+1\right)^{-2}$. 
Taking $\xi_0=r_s$,  the lens equation becomes \cite{bartel,zhao}
\beq
y=x-{4\rho_sr_s\over \Sigma_{cr}}{g(x)\over x},
\label{lens-eq:nfw}
\eeq
where $g(x)$ is defined by
\beq
g(x)=\ln{x\over 2} + {2\over \sqrt{1-x^2}}\tanh^{-1} \sqrt{{1-x\over 1+x}},
\label{g:nfw}
\eeq
for $x< 1$ and by
\beq
g(x)=\ln{x\over 2} + {2\over \sqrt{x^2-1}}\tan^{-1} \sqrt{{x-1\over x+1}}.
\eeq
for $x> 1$. The circular velocity is
\beq
v^2(r)={4\pi G\rho_sr_s^3\over r}\left(\ln(1+r/r_s)-{r/r_s\over 1+r/r_s}
\right).
\label{vel}
\eeq

%\subsection{truncated isothermal model}
%\paragraph{truncated isothermal model}

The truncated isothermal sphere (TIS) model is a particular 
solution of the Lane-Emden equation that results from the collapse and 
virialization of a top-hat density perturbation. \cite{tis} 
The mass profile is fit well by
\beq
\rho(r)=\rho_0\left({A\over (r/r_c)^2+a^2}-{B\over (r/r_c)^2+b^2}\right),
\eeq
where $(A,a^2,B,b^2)=(21.38,9.08,19.81,14.62)$. 
Shapiro et al. \cite{tis} found that this fitting formula is accurate within 
$3\%$ over $0 \leq r/r_c \siml 30$ for both the Einstein de Sitter model 
and low-density models $(\Omega_M \simg 0.3)$. 
The circular velocity is
\beq
v^2(r)=4\pi G\rho_0r_c^2\left(A-B-{aA\over\ x}\tan^{-1}{{x\over a}}+
{bB\over x}\tan^{-1}{{x\over b}} \right),
\eeq
where $x=r/r_c$.

%\subsection{observables, lens model parameters, and cosmological parameters}
%\paragraph{observables, lens model parameters, and cosmological parameters}

At this stage, it may be illuminating to estimate the sensitivity of the 
observable ($\theta_E$) to the parameters in lens model and 
the cosmological parameters. Using Eq.(\ref{th:iso}), for the case of an 
isothermal sphere with a core, the sensitivity is evaluated around
$r_c=0, \Omega_M=0.3, \Omega_{\Lambda}=0.7, w=-1$ as
\beq
{\delta\theta_E\over \theta_E}\simeq -
{\theta_c\over \theta_E}{\delta\theta_c\over \theta_c}+
2{\delta\sigma\over \sigma}-7.5\times 10^{-2}{\delta\Omega_M\over \Omega_M}+
2.9\times 10^{-2}{\delta w\over w},\label{relation}
\eeq
where $w$ is the equation of state of dark energy ($w=-1$ for the 
cosmological constant), and the source redshift and the lens redshift
are taken as $z_s=1.675$ and $z_l=0.39$, respectively, and a 
flat FRW model is assumed. 
%&\simeq& 1.8{\delta\rho_s\over \rho_s}+
%2.8{\delta\theta_s \over \theta_s}-1.1{\delta\Omega_M\over \Omega_M}+
%1.4{\delta w\over w}, \nonumber
%\eeqa
%where $\delta_c=\rho_s/\rho_{crit}$ and $\theta_s=r_s/D_L$,and $\rho_{crit}$
%is the critical density. 
It should be noted that in reality there should be contribution from 
the eccentricity of the lens profile added  to the above relation. 
Eq.(\ref{relation}) and its counterpart for the NFW profile 
(Eq.(\ref{relation:nfw})) indicate clearly that 
the lensing observable $\theta_E$ is primarily dependent on the 
lens model parameters, $\theta_c$ and $\sigma$,  and less dependent on 
the cosmological parameters, $\Omega_M$ and $w$. (This sensivity was 
noted previously in Ref.\cite{asada} for the case of an isothermal sphere.)
This is essentially because strong lensing is sensitive only to the mass 
inside the Einstein radius. 
For example, in order to put a constraint on $w$, one needs to measure 
the Einstein radius and  the velocity dispersion within ${\cal O}(1)\%$ 
accuracy and determine $\Omega_M$ within ${\cal O}(10)\%$ 
accuracy. The former requirement would be an observational 
challenge,\cite{fy} while the latter could be accomplished.\cite{haiman}
Another interesting observable is the location of radial arc,
 \cite{mh}\tocite{wnb} which depends on the angular gradient of 
the projected mass, as expressed by Eq.(\ref{critical}).

\vspace{10pt}
{\bf 3.Constraining the lens model by measurements of the velocity dispersion.}
%\subsection{case study: CL 0024+1654}
We now show that there exists a degeneracy of lens models to a certain 
extent with regard to the projected mass density and that this 
degeneracy persists even when measurements of the velocity dispersion
are included. We have in mind an arc or Einstein ring system, and 
thus one of the  observables is the critical radius. Of course, 
the length of arc is another important observable, \cite{length} but 
we do not consider it here.

As an illustration, we consider the well-known lensing system CL 0024+1654,
although our argument is not limited to cluster lenses. 
Bright multiple arcs were discovered in the cluster CL 0024+1654 at 
$z=0.39$ by Koo \cite{koo} photographically. 
Five arcs are clearly seen in the {\it HST} image (see Fig. 1 in
 Ref.~\citen{ctt}). 
The redshift of the source galaxy was recently determined
spectroscopically as $z_s=1.675$. \cite{broad} 
The distance to the arc from the center of the cluster is 
$\theta_E=34.6''$ \cite{wnb}, corresponding to $110 h^{-1}{\rm kpc}$ for 
the Einstein-de Sitter model. Here $h$ is the 
Hubble parameter in units of $100{\rm km/s/Mpc}$.

Tyson et al. \cite{tyson} attempted to construct a high-resolution mass 
map of the cluster CL 0024+1654 using the {\it Hubble Space Telescope}. 
They  found that the total mass profile within the arc radius 
is approximately represented by a power-law model, \cite{sef}
\beq
\Sigma(x)={K(1+\gamma x^2)\over (1+x^2)^{2-\gamma}},
\eeq
where $x=r/r_{\rm core}, K=7900\pm 100 h M_{\odot}{\rm pc}^{-2}, 
r_{\rm core}=35\pm 3 h^{-1} {\rm kpc}$ and $\gamma=0.57\pm 0.02$. 
They also noted that the asymmetry in the mass distribution inside 
the arcs for CL 0024+1654 is very small (less than 3$\%$).

Recently, however, Broadhurst et al. \cite{broad} have suggested that 
the mass profile of CL 0024+1654 is consistent with the NFW profile with 
$r_s\simeq 400h^{-1}{\rm kpc}$ and 
$\delta_c=\rho_s/\rho_{crit}\simeq 8000$. Here $\rho_{crit}$ 
is the critical density. 
Similarly to Eq.(\ref{relation}),  for the case of the NFW profile, using 
Eqs.(\ref{lens-eq:nfw}-\ref{g:nfw}), we obtain the following relation
for the above set of parameter values:
\beq
{\delta\theta_E\over \theta_E} \simeq 2.0{\delta\rho_s\over \rho_s}+
3.0{\delta r_s\over r_s}-0.23{\delta\Omega_M\over \Omega_M}+
0.19{\delta w\over w}.\label{relation:nfw}
\eeq 
Shapiro and Iliev found that the projected mass 
density profile obtained by Tyson et al. \cite{tyson} is fit well by 
that obtained from the TIS profile with 
$\rho_0\simeq 0.064 h^2M_{\odot}{\rm pc}^{-3}$ and 
$r_c\simeq 20h^{-1} {\rm kpc}$.\cite{shapiro} They also claimed that 
the mass profile obtained by Broadhurst et al. implies a velocity dispersion 
$(> 2230 {\rm km/s})$ that is much larger than the measured value.

\begin{figure}[htdp]
  \begin{center}
  \leavevmode\psfig{figure=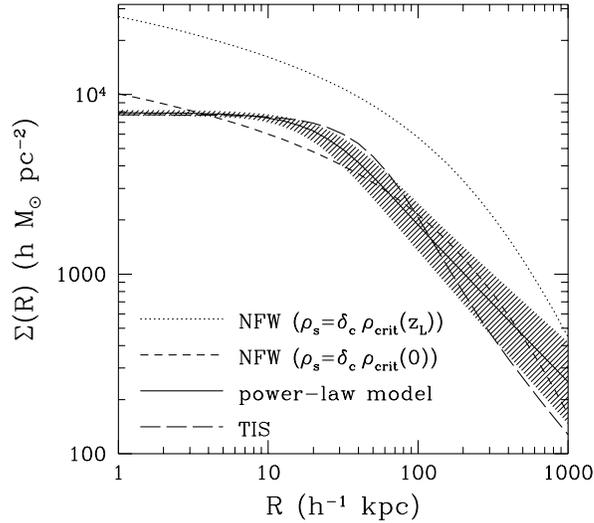,width=8cm}
  \end{center}
\caption{The projected mass density profile for CL 0024+1654.
The dotted line is the NFW profile with $\rho_s=\delta_c \rho_{crit}(z_L)$ 
used by Shapiro and Iliev, and the short-dashed line is the same with 
$\rho_s=\delta_c\rho_{crit}(0)$, where, $z_L=0.39$ and we assume the 
Einstein-de Sitter model.
The solid line the mass profile fitted by Tyson et al. with the 2$\sigma$ 
uncertainty by shades. 
The long-dashed line is the TIS profile fitted by Shapiro and Iliev. }
\end{figure}

In Fig. 1, we show the projected mass density profiles for these three models.
We assume the Einstein-de Sitter model and use two critical densities 
($\rho_{crit}(z=0)$ and $\rho_{crit}(z=z_L)$) for the fit with the NFW 
profile. We note that the angular resolution of the 
{\it Hubble Space Telescope} is $0.1''$, \cite{ctt} corresponding 
to $0.32 h^{-1}{\rm kpc}$. 
The shaded region is the two-sigma interval of the mass profile 
determined by Tyson et al. We assume that the parameters 
$(K,r_{\rm core},\gamma)$ are Gaussian-distributed with dispersions 
equal to the error bars. Within the uncertainties on the fitting
parameters, the power-law profile and the TIS profile look 
similar,\footnote{Using the Davidon-Fletcher-Powell method, we 
independently fit Tyson's profile with the TIS model and found that 
Tyson's profile is also fit well (within $100 {\rm kpc}$) by the TIS 
profile with $\rho_0\simeq 0.0837 h^2M_{\odot}{\rm pc}^{-3}$ and 
$r_c\simeq 15.4h^{-1} {\rm kpc}$.} while the NFW profile with 
$\rho_{crit}(z=0)$ deviates slightly from the power-law profile 

As is clear from Fig. 1, the problem with the NFW mass profile fitted by 
Broadhurst et al. may not result from a problem with the NFW model itself 
but a problem with the definition of $\rho_{crit}$ 
used in the analysis. In the original NFW fit of the cold dark matter halo 
profile, $\rho_{crit}$ in the relation $\rho_s=\delta_c \rho_{crit}$
should be evaluated at the redshift of the object. However, within a 
fitting model, it is not necessary to do this, and we can treat $\rho_s$ 
just as a free parameter of the model. If we use $\rho_{crit}(z=z_L)$ as 
Shapiro and Iliev \cite{shapiro} did, then the projected mass density 
is much higher than that obtained from the Tyson's fit of the data from 
the beginning.\footnote{In fact, T. Broadhurst informed us that they used 
$\rho_{crit}(z=0)$ to normalize $\rho_s$.}
%In fact, if we use $\rho_{crit}(z=z_L)$ as Shapiro and Iliev 
%\shortcite{shapiro} did, then the projected mass density 
%is much higher than the Tyson's fit of the data from the beginning.

As is evident from the lens equation Eq.(\ref{lens}), 
gravitational lensing provides information regarding the projected 
two-dimensional mass density. Therefore one may wonder if the degeneracy 
of the lens model could be lifted by combining with three-dimensional data, 
for example, the velocity dispersion. However, we believe that this is 
unlikely. The reason is the following. {}From Eq.(\ref{vel:iso}) the 
sensitivity of the velocity dispersion on the parameters of the lens model 
is evaluated as 
\beq
{\delta v\over v}\simeq  -{\pi \delta r_c\over 4r}
+{\delta \sigma\over \sigma}.
\eeq
Comparing with Eq.(\ref{relation}), this indicates that the sensivity 
of the velocity dispersion is less dependent than the sensitivity of
$\theta_E$ on the parameters of the lens model. The size of the system
at which the velocity dispersion is measured is larger than the 
Einstein radius. 

\begin{figure}[htdp]
  \begin{center}
  \leavevmode\psfig{figure=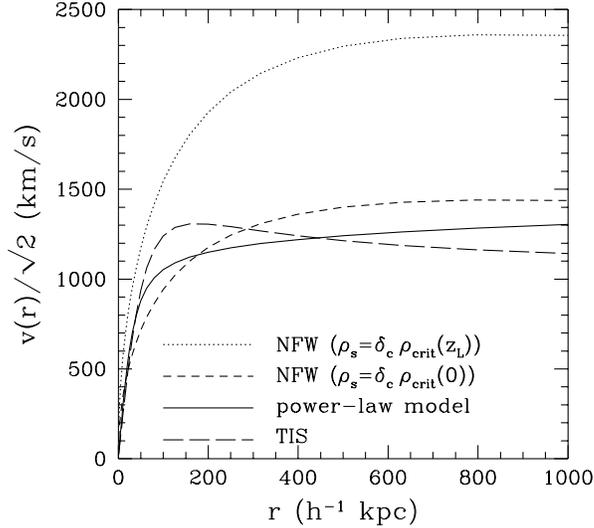,width=8cm}
  \end{center}
\caption{
The circular velocity divided by $\sqrt{2}$, which coincides 
with the velocity dispersion for a singular isothermal lens, 
for the NFW model with $\rho_s=\delta_c\rho_{crit}(z_L)$ (the dotted line), 
the same with $\rho_s=\delta_c\rho_{crit}(0)$ (the short-dashed line)  
the  power-law model (the solid line), 
and for the TIS profile (the long-dashed line).
Here $z_L=0.39$ and we assume the Einstein-de Sitter model.
The average velocity dispersion of CL 0024+1654 is measured to be 1150 km/s
within a radius $r \simeq 600 h^{-1}$ km/s to an accuracy $\pm$ 100 km/s.
}
\end{figure}

In Fig. 2, we plot the circular velocity profiles divided by $\sqrt{2}$
for each mass model, although it is known that  $v(r)/\sqrt{2}$ exactly 
coincides with the velocity dispersion only for a singular isothermal
lens. The velocity profile of the power-law model was calculated using 
the Abel integral.\cite{bt} As suggested by Fig. 1, the NFW fit with 
$\rho_{crit}(z=z_L)$ predicts a velocity profile that is much higher than 
the measured value, in accordance with the claim of Shapiro and Iliev 
\cite{shapiro}. We find that the velocity profiles
are all similar. The average velocity dispersion of CL 0024+1654 has
been  measured to be $1150 {\rm km/s}$ within a radius 
$r\simeq 600 h^{-1} {\rm kpc}$, 
based on $107$ galaxy redshifts \cite{dressler}, to an accuracy of 
roughly $\pm 100{\rm km s^{-1}}$. For $33$ galaxy redshifts, it has also 
been measured to be $1390{\rm km/s}$ \cite{smail}.

%\subsection{case study: PG 1115+080}

\vspace{10pt}
%\section{summary}
{\bf 4.Summary.}
We have examined the relation between lensing observables and  
model parameters (the lens model and the cosmological parameters). 
We have found that observables are primarily dependent on the lens model
parameters and have assessed the accuracy required to determine 
cosmological parameters [as expressed by Eq.(\ref{relation}) and 
Eq.(\ref{relation:nfw})]. 
It is the surface density of a lens that can be determined from 
the observations of gravitational lensing, and therefore there exists a 
degeneracy among lens models given the observational uncertainties.
This degeneracy cannot be broken even by combining measurements of the 
velocity dispersion. We also have identified possible source of the 
problem of fitting the averaged mass profile of CL 0024+1654 
with the NFW profile. The reconstruction of the mass profile 
using detailed shear maps with weak lensing observations \cite{weak,ss} 
may provide more accurate information regarding the mass profile of
lenses. In any case, it is not possible to put meaningful constraints on 
cosmological parameters using gravitational lensing until we obtain more 
complete understanding of the lens model.

\section*{Acknowledgements}
We would like to thank Tom Broadhurst for useful correspondence 
and Toshifumi Futamase for useful comments.  
This work was supported in part by a 
Grant-in-Aid for Scientific Research (No.13740154) from the Japan Society for 
the Promotion of Science.

\end{document}